\documentclass{PoS}
\usepackage{graphics}

\title{Forces between static-light mesons}

\ShortTitle{Forces between static-light mesons}

\author{\speaker{Marc Wagner}\\
        Humboldt-Universit\"at zu Berlin, Institut f\"ur Physik, Newtonstra{\ss}e 15, D-12489 Berlin, Germany\\
        E-mail: \email{mcwagner@physik.hu-berlin.de}}

\abstract{
The isospin, spin and parity dependent potential of a pair of static-light mesons is computed using Wilson twisted mass lattice QCD with two flavors of degenerate dynamical quarks. From the results a simple rule can be deduced stating, which isospin, spin and parity combinations correspond to attractive and which to repulsive forces.
}

\FullConference{The XXVIII International Symposium on Lattice Field Theory, Lattice2010\\
		June 14-19, 2010\\
		Villasimius, Italy}

\newcommand{\ltapprox}{\raisebox{-0.5ex}{$\,\stackrel{<}{\scriptstyle\sim}\,$}}

\begin{document}


\section{Introduction}

Lattice computations of the potential of a pair of static-light mesons (in the following also referred to as $B$ mesons) are of interest, because they constitute first principles determinations of a hadronic force. Until now interactions between static-light mesons have exclusively been studied in the quenched approximation \cite{Michael:1999nq,Detmold:2007wk}. Here I report on the status of an investigation with two flavors of dynamical Wilson twisted mass quarks. Forces are not only studied between the lightest static-light mesons (denoted by $S$), but also first excitations are taken into account (denoted by $P_-$). Note that there is another ongoing study of static-light meson interactions with dynamical quarks, which has also been reported during this conference \cite{BH2010}.


\section{Trial states and quantum numbers}


\subsection{\label{SEC001}Static-light mesons}

Here I consider static-light mesons, which are made from a static antiquark $\bar{Q}$ and a light quark $\psi \in \{ u \, , \, d \}$. Consequently, isospin $I = 1/2$ and $I_z \in \{ -1/2 \, , \, +1/2 \}$. Since there are no interactions involving the static quark spin, it is appropriate to classify static-light mesons by the angular momentum of their light degrees of freedom $j$. I do not consider non-trivial gluonic excitations, hence $j = 1/2$ and $j_z = \{ -1/2 \, , \, +1/2 \}$, which is the spin of the light $u/d$ quark. Parity is also a quantum number, $\mathcal{P} \in \{ + \, , \, - \}$.

The lightest static-light meson has quantum numbers $j^\mathcal{P} = (1/2)^-$ (denoted by $S$). The first excitation, which is $\approx 400 \, \textrm{MeV}$ heavier, has quantum numbers $j^\mathcal{P} = (1/2)^+$ (denoted by $P_-$). Examples of corresponding static-light meson trial states are $\bar{Q} \gamma_5 \psi | \Omega \rangle$ and $\bar{Q} \gamma_j \psi | \Omega \rangle$ for $S$ mesons and $\bar{Q} \psi | \Omega \rangle$ and $\bar{Q} \gamma_j \gamma_5 \psi | \Omega \rangle$ for $P_-$ mesons respectively.

For a more detailed discussion of static-light mesons I refer to \cite{Jansen:2008si,:2010iv}.


\subsection{\label{SEC002}$B B$ systems}

The aim of this work is to determine the potential of a pair of $B$ mesons as a function of their separation $R$ (without loss of generality I choose the axis of separation to be the $z$ axis). To this end one has to compute the energy of eigenstates of the Hamiltonian containing two static antiquarks $\bar{Q}(\mathbf{r}_1)$ and $\bar{Q}(\mathbf{r}_2)$, $\mathbf{r}_1 = (0,0,-R/2)$ and $\mathbf{r}_2 = (0,0,+R/2)$, which define the positions of the two $B$ mesons, and which will be surrounded by light quarks and gluons.

These $B B$ states are characterized by several quantum numbers. Since there are two light $u/d$ valence quarks, isospin $I \in \{ 0 \, , \, 1 \}$ and $I_z \in \{ -1 \, , \, 0 \, , \, +1 \}$. Due to the separation of the static antiquarks along the $z$ axis, rotational symmetry is restricted to rotations around this axis. Consequently, states can be classified by the $z$ component of total angular momentum. However, as already mentioned in section~\ref{SEC001} there are no interactions involving the static quark spin. Therefore, it is appropriate to label $B B$ states by the $z$ component of the angular momentum of the light degrees of freedom $j_z \in \{ -1 \, , \, 0 \, , \, +1 \}$. Parity is also a symmetry and, therefore, a quantum number, $\mathcal{P} \in \{ + \, , \, - \}$. For states with $j_z = 0$ there is an additional symmetry, reflection along an axis perpendicular to the axis of separation (without loss of generality I choose the $x$ axis). The corresponding quantum number is $\mathcal{P}_x \in \{ + \, , \, - \}$. When using $|j_z|$ instead of $j_z$, $\mathcal{P}_x$ is a quantum number for all states. To summarize, $B B$ states can be characterized by the following five quantum numbers: $(I , I_z , |j_z| , \mathcal{P} , \mathcal{P}_x)$.

I use $B B$ trial states
\begin{eqnarray}
\label{EQN001} (\mathcal{C} \Gamma)_{A B} \Big(\bar{Q}_C(\mathbf{r}_1) \psi_A^{(1)}(\mathbf{r}_1)\Big) \Big(\bar{Q}_C(\mathbf{r}_2) \psi_B^{(2)}(\mathbf{r}_2)\Big) | \Omega \rangle ,
\end{eqnarray}
where the lower indices $A$, $B$ and $C$ denote spinor indices, $\mathcal{C} = \gamma_0 \gamma_2$ is the charge conjugation matrix and $\Gamma$ is a combination of $\gamma$ matrices. Note that it is essential to couple the light degrees of freedom of both mesons in spinor space, because these degrees of freedom determine the quantum number $|j_z|$. Proceeding in a naive way by coupling light and static degrees of freedom in both $B$ mesons separately will not result in a well defined angular momentum $|j_z|$ and, therefore, will mix different sectors. To obtain $I = 0$, the flavors of the light quarks have to be chosen according to $\psi^{(1)} \psi^{(2)} = u d - d u$, while for $I = 1$ three possibilities exist, $\psi^{(1)} \psi^{(2)} \in \{ u u \, , \, d d \, , \, ud + d u \}$. $B B$ trial states are collected in Table~\ref{TAB001} together with their quantum numbers.


\begin{table}[htb]
\begin{center}

\begin{tabular}{|c|c||c|c||c|c||c|c|}
\hline
\multicolumn{2}{|c||}{\vspace{-0.40cm}} & \multicolumn{2}{c||}{} & \multicolumn{2}{c||}{} & \multicolumn{2}{c|}{} \\
\multicolumn{2}{|c||}{} & \multicolumn{2}{c||}{$\psi^{(1)} \psi^{(2)} = u d - d u$} & \multicolumn{2}{c||}{$\psi^{(1)} \psi^{(2)} = u d + d u$} & \multicolumn{2}{c|}{$\psi^{(1)} \psi^{(2)} \in \{ u u \, , \, d d \}$} \\
\multicolumn{2}{|c||}{\vspace{-0.40cm}} & \multicolumn{2}{c||}{} & \multicolumn{2}{c||}{} & \multicolumn{2}{c|}{} \\
\hline
 & & & & & & & \vspace{-0.40cm} \\
$\Gamma$ & $|j_z|$ & $\mathcal{P}$, $\mathcal{P}_x$ & result & $\mathcal{P}$, $\mathcal{P}_x$ & result & $\mathcal{P}$, $\mathcal{P}_x$ & result \\
 & & & & & & & \vspace{-0.40cm} \\
\hline
 & & & & & & & \vspace{-0.40cm} \\
$\gamma_5$                   & $0$ & $-$, $+$ & A, SS & $+$, $+$ & R, SS & $+$, $+$ & R, SS \\
$\gamma_0 \gamma_5$          & $0$ & $-$, $+$ & A, SS & $+$, $+$ & R, SS & $+$, $+$ & R, SS \\
$1$                          & $0$ & $+$, $-$ & A, SP & $-$, $-$ & R, SP & $-$, $-$ & R, SP \\
$\gamma_0$                   & $0$ & $-$, $-$ & R, SP & $+$, $-$ & A, SP & $+$, $-$ & A, SP \\
$\gamma_3$                   & $0$ & $+$, $-$ & R, SS & $-$, $-$ & A, SS & $-$, $-$ & A, SS \\
$\gamma_0 \gamma_3$          & $0$ & $+$, $-$ & R, SS & $-$, $-$ & A, SS & $-$, $-$ & A, SS \\
$\gamma_3 \gamma_5$          & $0$ & $+$, $+$ & A, SP & $-$, $+$ & R, SP & $-$, $+$ & R, SP \\
$\gamma_0 \gamma_3 \gamma_5$ & $0$ & $-$, $+$ & R, SP & $+$, $+$ & A, SP & $+$, $+$ & A, SP \\
 & & & & & & & \vspace{-0.40cm} \\
\hline
 & & & & & & & \vspace{-0.40cm} \\
$\gamma_{1/2}$                   & $1$ & $+$, $\pm$ & R, SS & $-$, $\pm$ & A, SS & $-$, $\pm$ & A, SS \\
$\gamma_0 \gamma_{1/2}$          & $1$ & $+$, $\pm$ & R, SS & $-$, $\pm$ & A, SS & $-$, $\pm$ & A, SS \\
$\gamma_{1/2} \gamma_5$          & $1$ & $+$, $\mp$ & A, SP & $-$, $\mp$ & R, SP & $-$, $\mp$ & R, SP \\
$\gamma_0 \gamma_{1/2} \gamma_5$ & $1$ & $-$, $\mp$ & R, SP & $+$, $\mp$ & A, SP & $+$, $\mp$ & A, SP\vspace{-0.40cm} \\
 & & & & & & & \\
\hline
\end{tabular}

\caption{\label{TAB001}quantum numbers of $B B$ trial states; due to explicit isospin breaking, $(I = 1 , I_z = 0)$ and $(I = 1 , I_z = \pm 1)$ states are not degenerate in twisted mass lattice QCD (cf.\ section~3) and, therefore, listed separately; ``result'' characterizes the shapes of the numerically computed $B B$ potentials (A: attractive potential; R: repulsive potential; SS: lower asymptotic value $2 m(S)$; SP: higher asymptotic value $m(S) + m(P_-)$; cf.\ section~4).}

\end{center}
\end{table}



\section{Lattice setup}

I use $24^3 \times 48$ gauge field configurations generated by the European Twisted Mass Collaboration (ETMC). The fermion action is $N_f = 2$ Wilson twisted mass,
\begin{eqnarray}
S_\mathrm{F}[\chi,\bar{\chi},U] \ \ = \ a^4 \sum_x \bar{\chi}(x) \Big(D_\mathrm{W} + i\mu_\mathrm{q}\gamma_5\tau_3\Big) \chi(x)
\end{eqnarray}
\cite{Frezzotti:2000nk,Frezzotti:2003ni}, where $D_\mathrm{W}$ is the standard Wilson Dirac operator and $\chi = (\chi^{(u)} , \chi^{(d)})$ is the light quark doublet in the so-called twisted basis. In the continuum the twisted basis is related to the physical basis by the twist rotation $\psi = e^{i \gamma_5 \tau_3 \omega / 2} \chi$, where $\omega$ is the twist angle. $\omega$ has been tuned to maximal twist, i.e.\ $\omega = \pi / 2$, where static-light mass differences are automatically $\mathcal{O}(a)$ improved. The gauge action is tree-level Symanzik improved \cite{Weisz:1982zw}. I use $\beta = 3.9$ and $\mu_\mathrm{q} = 0.0040$ corresponding to a lattice spacing $a = 0.079(3) \, \textrm{fm}$ and a pion mass $m_\mathrm{PS} = 340(13) \, \textrm{MeV}$ \cite{Baron:2009wt}. For details regarding these gauge field configurations I refer to \cite{Boucaud:2007uk,Boucaud:2008xu}.



In twisted mass lattice QCD at finite lattice spacing SU(2) isospin is explicitely broken to U(1), i.e.\ $I_z$ is still a quantum number, but $I$ is not. Moreover, parity $\mathcal{P}$ has to be replaced by twisted mass parity $\mathcal{P}^{(\textrm{\scriptsize tm})}$, which is parity combined with light flavor exchange. The consequence is that twisted mass $B B$ sectors are either labeled by $(I_z , |j_z| , \mathcal{P}^{(\textrm{\scriptsize tm})} \mathcal{P}_x^{(\textrm{\scriptsize tm})})$ for $I_z = \pm 1$ or by $(I_z , |j_z| , \mathcal{P}^{(\textrm{\scriptsize tm})} , \mathcal{P}_x^{(\textrm{\scriptsize tm})})$ for $I_z = 0$. A comparison with the set of quantum numbers discussed in section~\ref{SEC002} shows that in the twisted mass formalism there are only half as many $B B$ sectors as in QCD, i.e.\ QCD $B B$ sectors are pairwise combined. Nevertheless, it is possible to unambiguously interpret states obtained from twisted mass correlation functions in terms of QCD quantum numbers. The method has successfully been applied in the context of static-light mesons \cite{Blossier:2009vy} and is explained in detail for kaons and $D$ mesons in \cite{Baron:2010th}. For a detailed discussion of twisted mass symmetries in the context of $B B$ systems I refer to an upcoming publication \cite{MW2010}.

When computing correlation functions, I use several techniques to improve the signal quality including operator optimization by means of APE and Gaussian smearing and stochastic propagators combined with timeslice dilution. These techniques are very similar to those used in a recent study of the static-light meson spectrum \cite{Jansen:2008si,:2010iv} and will also be explained in detail in \cite{MW2010}.

\begin{figure}[htb]
\begin{center}
\input{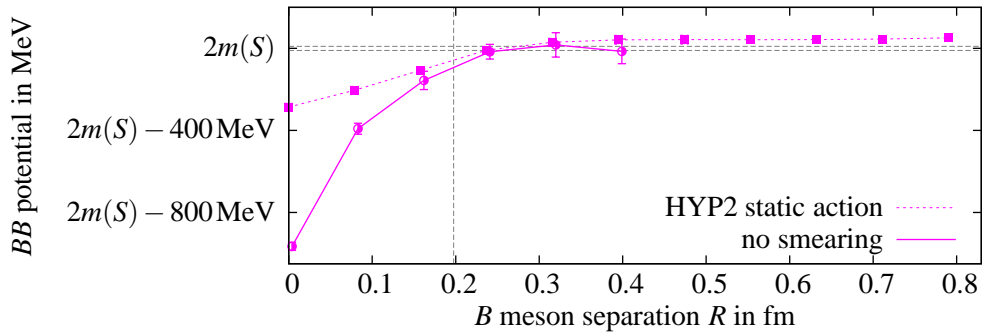}
\caption{\label{FIG002}the $B B$ potential corresponding to $\psi^{(1)} \psi^{(2)} = uu$, $\Gamma = \gamma_3$ computed with unsmeared links and with the HYP2 static action.}
\end{center}
\end{figure}

In contrast to spectrum calculations for static-light mesons \cite{Jansen:2008si,:2010iv} and static-light baryons \cite{Wagner:2010hj}, where we have always used the HYP2 static action, I perform computations both with the HYP2 static action and with unsmeared links representing the world lines of the static antiquarks. In particular for small $\bar{Q} \bar{Q}$ separations $R \ltapprox 2 a$ ultraviolet fluctuations are important, which are, however, filtered out, when using HYP smeared links. The effect of HYP smearing is shown in Figure~\ref{FIG002}. For all results presented in the following potential values corresponding to $R \leq 2 a$ have been computed by means of unsmeared links, while for larger separations HYP smearing has been applied to improve the signal-to-noise ratio.

\section{Numerical results}

\begin{figure}[htb]
\begin{center}
\input{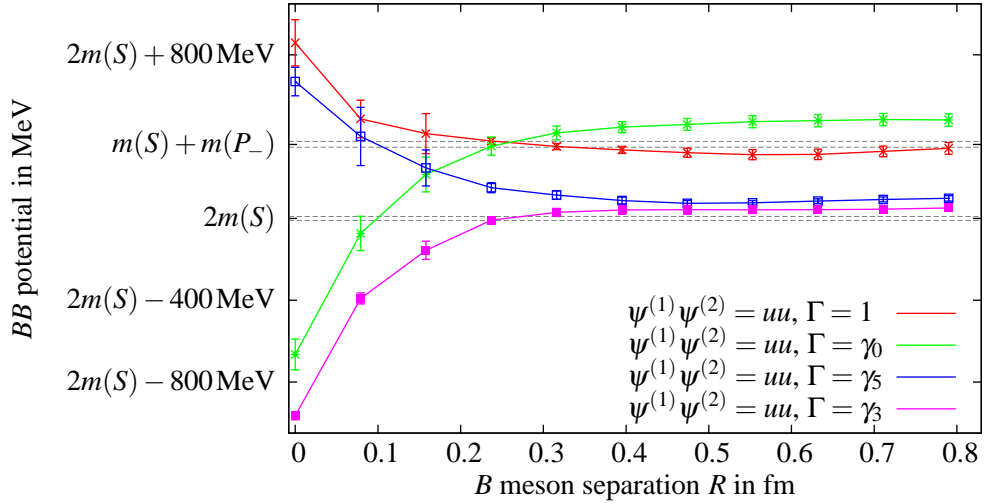}
\caption{\label{FIG001}examples of $B B$ potentials as functions of the separation $R$.}
\end{center}
\end{figure}

The $B B$ potentials presented and discussed in the following have been obtained by fitting constants to effective mass plateaus obtained from temporal correlation functions of trial states (\ref{EQN001}). In twisted mass lattice QCD there are 24 independent $I_z = 0$ trial states (i.e.\ trial states not related by symmetries) and 12 independent $I_z = \pm 1$ trial states, i.e.\ 36 resulting  potentials, which are not related by symmetries (cf.\ Table~\ref{TAB001}). Some of these potentials are quite similar, while others are not. In total there are four significantly different types of potentials: two of them are attractive, the other two are repulsive; two have have asymptotic values for large separations $R$, which are larger by around $400 \, \textrm{MeV}$ compared to the other two (cf.\ the ``result'' columns of Table~\ref{TAB001}). For each of the four types an example is plotted in Figure~\ref{FIG001}.

To understand the asymptotic behavior, it is convenient to express the $B B$ creation operators appearing in (\ref{EQN001}) in terms of static-light meson creation operators. For the potentials shown in Figure~\ref{FIG001} one finds after some linear algebra
\begin{eqnarray}
\nonumber & & \hspace{-0.7cm} (\mathcal{C} 1)_{A B} \Big(\bar{Q}_C(\mathbf{r}_1) u_A(\mathbf{r}_1)\Big) \Big(\bar{Q}_C(\mathbf{r}_2) u_B(\mathbf{r}_2)\Big) \ \ = \\
\label{EQN_g5} & & = \ \ -S_\uparrow(\mathbf{r}_1) P_{- \downarrow}(\mathbf{r}_2) + S_\downarrow(\mathbf{r}_1) P_{- \uparrow}(\mathbf{r}_2) - P_{- \uparrow}(\mathbf{r}_1) S_\downarrow(\mathbf{r}_2) + P_{- \downarrow}(\mathbf{r}_1) S_\uparrow(\mathbf{r}_2) \\
\nonumber & & \hspace{-0.7cm} (\mathcal{C} \gamma_0)_{A B} \Big(\bar{Q}_C(\mathbf{r}_1) u_A(\mathbf{r}_1)\Big) \Big(\bar{Q}_C(\mathbf{r}_2) u_B(\mathbf{r}_2)\Big) \ \ = \\
\label{EQN_g0} & & = \ \ -S_\uparrow(\mathbf{r}_1) P_{- \downarrow}(\mathbf{r}_2) + S_\downarrow(\mathbf{r}_1) P_{- \uparrow}(\mathbf{r}_2) + P_{- \uparrow}(\mathbf{r}_1) S_\downarrow(\mathbf{r}_2) - P_{- \downarrow}(\mathbf{r}_1) S_\uparrow(\mathbf{r}_2) \\
\nonumber & & \hspace{-0.7cm} (\mathcal{C} \gamma_5)_{A B} \Big(\bar{Q}_C(\mathbf{r}_1) u_A(\mathbf{r}_1)\Big) \Big(\bar{Q}_C(\mathbf{r}_2) u_B(\mathbf{r}_2)\Big) \ \ = \\
\label{EQN_1} & & = \ \ -S_\uparrow(\mathbf{r}_1) S_\downarrow(\mathbf{r}_2) + S_\downarrow(\mathbf{r}_1) S_\uparrow(\mathbf{r}_2) - P_{- \uparrow}(\mathbf{r}_1) P_{- \downarrow}(\mathbf{r}_2) + P_{- \downarrow}(\mathbf{r}_1) P_{- \uparrow}(\mathbf{r}_2) \\
\nonumber & & \hspace{-0.7cm} (\mathcal{C} \gamma_3)_{A B} \Big(\bar{Q}_C(\mathbf{r}_1) u_A(\mathbf{r}_1)\Big) \Big(\bar{Q}_C(\mathbf{r}_2) u_B(\mathbf{r}_2)\Big) \ \ = \\
\label{EQN_g3} & & = \ \ -i S_\uparrow(\mathbf{r}_1) S_\downarrow(\mathbf{r}_2) -i S_\downarrow(\mathbf{r}_1) S_\uparrow(\mathbf{r}_2) +i P_{- \uparrow}(\mathbf{r}_1) P_{- \downarrow}(\mathbf{r}_2) +i P_{- \downarrow}(\mathbf{r}_1) P_{- \uparrow}(\mathbf{r}_2) .
\end{eqnarray}
At large separations $R$ the $B B$ potentials are expected to approach the sum of the masses of the two individual $B$ mesons. When considering (\ref{EQN_g5}) to (\ref{EQN_g3}) and Figure~\ref{FIG001}, one can see that the two potentials with the lower asymptotic value ($\psi^{(1)} \psi^{(2)} = uu$, $\Gamma = \gamma_5$ and $\psi^{(1)} \psi^{(2)} = uu$, $\Gamma = \gamma_3$) contain $S S$ combinations. These are significantly lighter than the also present $P_- P_-$ combinations and should, therefore, dominate the correlation functions and effective masses at large temporal separations. The asymptotic value of the corresponding potentials should be around $2 m(S)$, which is the case. In contrast to that the other two potentials with the higher asymptotic value \\ ($\psi^{(1)} \psi^{(2)} = uu$, $\Gamma = 1$ and $\psi^{(1)} \psi^{(2)} = uu$, $\Gamma = \gamma_0$) exclusively contain $S P_-$ combinations. Their asymptotic value is expected at around $m(S) + m(P_-)$, which is also reflected by Figure~\ref{FIG001}.

This expansion of $B B$ creation operators in terms of static-light meson creation operators also provides an explanation, why potentials computed with different operators, but which have identical quantum numbers, are of different type. An example is given by $\psi^{(1)} \psi^{(2)} = uu$, $\Gamma = \gamma_3$ and $\psi^{(1)} \psi^{(2)} = uu$, $\Gamma = 1$, both having quantum numbers $(I = 1, I_z = +1, |j_z| = 0, \mathcal{P} = -, \mathcal{P}_x = -)$. The $\Gamma = \gamma_3$ potential is attractive with an asymptotic value at around $2 m(S)$, while the $\Gamma = 1$ potential is repulsive with an asymptotic value at around $m(S) + m(P_-)$. From (\ref{EQN_g5}) and (\ref{EQN_g3}) one can read off that the static-light meson content is essentially ``orthogonal'': the $\Gamma = \gamma_3$ operator contains $S S$ and $P_- P_-$ combinations, whereas the $\Gamma = 1$ operator is exclusively made from $S P_-$ combinations. While the corresponding $\Gamma = \gamma_3$ correlator yields the ground state in the $(I = 1, I_z = +1, |j_z| = 0, \mathcal{P} = -, \mathcal{P}_x = -)$ sector, which closely resembles a pair of $S$ mesons, the $\Gamma = 1$ operator mainly excites the first excitation, which is similar to an $S P_-$ combination. The generated ground state overlap is, therefore, rather small and, consequently, very large temporal separations would be needed to extract the ground state potential. Presumably, the potential corresponding to the $\Gamma = 1$ operator has a small ground state contribution, which contaminates the first excited state potential. This is supported by the observation that the asymptotic value of the $\Gamma = 1$ potential is slightly lower than $m(S) + m(P_-)$. For a clean extraction of this first excited state an analysis of a $2 \times 2$ correlation matrix is needed. 

From the 36 independent potentials one can also deduce a rule stating, whether a $B B$ potential is attractive or repulsive. The rule is quite simple. \\
%
%
\textbf{A }$B B$\textbf{ potential is attractive, if the trial state is symmetric under meson exchange, repulsive, if the trial state is antisymmetric under meson exchange.} \\
Here meson exchange means exchange of flavor, spin and parity. One can easily verify this rule for the examples discussed above: the operators (\ref{EQN_g0}) and (\ref{EQN_g3}) are symmetric under meson exchange and give rise to attractive potentials, while the operators (\ref{EQN_g5}) and (\ref{EQN_1}) are antisymmetric under meson exchange and yield repulsive potentials. This more general rule is in agreement to what has been observed in quenched $B B$ computations for $S S$ potentials \cite{Michael:1999nq,Detmold:2007wk}.


\section{Conclusions}

I have presented results of an ongoing computation of $B B$ potentials. Various channels characterized by the quantum numbers $(I , I_z , |j_z| , \mathcal{P} , \mathcal{P}_x)$ have been investigated. The computations have been performed with dynamical, rather light quark masses ($m_\mathrm{PS} \approx 340 \, \textrm{MeV}$). The results have been interpreted in terms of individual $S$ and $P_-$ mesons. A simple rule has been established stating, whether a $B B$ potential is attractive or repulsive.

The statistical accuracy of the correlation functions needs to be improved. $B B$ systems are rather heavy and, hence, effective masses are quickly lost in noise. At the present level of statistics slight contamination from excited states cannot be excluded. To this end contractions are ongoing.

Future plans include studying the light quark mass dependence, the continuum limit and finite volume effects. Moreover, also $B B_s$ and $B_s B_s$ potentials could be computed. To treat the $s$ quark as a fully dynamical quark, such computations should be performed on $N_f = 2+1+1$ flavor gauge field configurations currently produced by ETMC \cite{Baron:2010bv}. It would also be interesting to supplement the lattice computation by a perturbative calculation of $B B$ potentials at small separations $R \ltapprox 2$. Finally, one could use the obtained $B B$ potentials as input for phenomenological considerations to answer e.g.\ the question, whether two $B$ mesons are able to form a bound state.


\begin{acknowledgments}

I acknowledge useful discussions with Pedro Bicudo, William Detmold, Rudolf Faustov, \\ Roberto Frezzotti, Vladimir Galkin, Chris Michael and Attila Nagy. This work has been supported in part by the DFG Sonderforschungsbereich TR9 Computergest\"utzte The\-o\-re\-tische Teilchenphysik.

\end{acknowledgments}



\end{document}